# Extreme Low Thermal Conductivity in Nanoscale 3D Si Phononic Crystal with Spherical Pores

Lina Yang[1], Nuo Yang[2*], and Baowen Li[1,2,3,4*]

[1]Department of Physics and Centre for Computational Science and Engineering, National University of Singapore, Singapore 117542, Republic of Singapore

[2]Center for Phononics and Thermal Energy Science, School of Physical Science and Engineering, Tongji University, 200092 Shanghai, People's Republic of China

[3]Graphene Research Center, National University of Singapore, Singapore 117542, Republic of Singapore

[4]NUS Graduate School for Integrative Sciences and Engineering, National University of Singapore, Singapore 117456, Republic of Singapore

*Correspondence and requests for materials should be addressed to: N.Y. (imyangnuo@gmail.com) and B.L. (phylibw@nus.edu.sg)

**Thermoelectric material provides a high hope for converting harmful and useless heat into useful energy – electricity. It can also be used as solid-state Peltier coolers in integrated circuits –an outstanding challenge for electronic engineers. The desire for a high efficient thermoelectric material has never been so keen. Although many works have been done, we are still far from having a recipe for thermoelectric materials. Nano-structuring provides an effective way to increase figure of merit (ZT) by reducing the thermal conductivity without affecting electronic property.[1] Here, we propose a novel nanoscale three-dimensional (3D) Si phononic crystal (PnC) with spherical pores, which can reduce the thermal conductivity of bulk Si by a factor up to *10,000 times* at room temperature. The extreme-low thermal conductivity could lead to a larger value of ZT than unity. The thermal conductivity changes little when temperature increases from room temperature to 1100 K. The phonon participation ratio spectra show there are more phonon localizations as the porosity of PnC increases.**

Compared with other thermoelectric materials, Si nanostructures are low-cost, environment friendly and widely used in semiconductor industry.[2,3-5] Similar to photonic crystals, PnCs are constructed by a periodic array of scattering inclusions distributed in a host material. Since the period length decreases to nanometers, PnCs can affect the transport of terahertz lattice vibrations - phonons. Due to its periodic change of the density and /or elastic constants, PnCs exhibit phononic band gaps.[5] This remarkable property is very different from those of traditional materials and can be engineered to achieve new functionalities, such as modulating thermal conductivity.

A successful methodology has been reported to implement two-dimensional (2D) PnC geometry in Si,[6,7] which is compatible with standard CMOS fabrication and mass produced.

Experimental results demonstrated that 2D Si PnCs exhibited thermal conductivity as low as 2 W/m-K, meanwhile the values of electrical conductivity were comparable to those for bulk Si thin film.[6] Besides 2D PnCs, Gillet et. al predicted that the thermal conductivity of 3D PnC of Ge quantum-dots (QDs) in Si could be reduced by several orders of magnitude compared with bulk Si, using Boltzmann transport equation (BTE) method.[8] Our previous work showed 3D isotopic Si PnC could significantly reduce the thermal conductivity of bulk Si at high temperature (1000 K) by EMD simulations.[5] The reduction of thermal conductivity is due to a reduction of group velocities, phonon localizations and band gaps. Recently, Ma et. al measured that thermal conductivity of nanoscale 3D Si PnC is below 10 W/m-K.[9] Similar to 3D Si PnC structure, nanoporous Si[10,11] can also significantly reduce lattice thermal conductivity with specific choices of the pore size and spacing. On the other hand, Si PnCs and nanoporous Si could preserve their electrical properties with little degradation.[6,10] Consequently, the reduction of thermal conductivity in nanoscale Si PnCs and nanoporous Si could lead to a larger ZT than unity.

The 3D Si PnC is constructed by periodic arrangement of nanoscale supercell constructed from a cubic cell with a spherical pore (shown in Fig.1). The centers of cubic and spherical pore are overlapped. The period length of PnC is the distance between centers of two nearest supercell. The porosity is defined as the ratio of number of removed atoms in pore to the total number of atoms in a cubic Si cell.

To compare the thermal conductivity of 3D PnC with bulk Si, we calculate the thermal conductivity of bulk Si at 300 K as $170 \pm 16$ W/m-K by EMD method, where the volume of the simulation cell of bulk Si is $12 \times 12 \times 12$ units[3] (1 unit is 0.543 nm) with periodic boundary condition applied in three direction. Our result is comparable to Henry et al.'s results of EMD simulation, 160 W/m-K.[12] The experimental value of thermal conductivity of bulk Si at 300 K[13] is around 156 W/m-K, which means the thermal conductivity obtained from molecular dynamic (MD) method is a bit overestimated. This non-coincidence with measurement value is caused by both the inaccuracies of semi-empirical potentials and the impurity of the sample in measurements. However, this non-coincidence has little effect on comparing MD results with the same potential parameters.

We calculated the thermal conductivity of 3D PnCs with different porosity at room temperature. In all of the following simulations of PnCs, we use 16 units as the side length of simulation cell which is large enough to overcome the finite size effect (Finite size effect in Supporting Information I) and 8 units as period length. As shown in Fig. 2(a), the thermal conductivity rapidly decreases as the porosity increases. When the porosity is 30%, the thermal conductivity is 1.67 W/m-K. Moreover, when porosity increases to 90%, the thermal conductivity is decreased to 0.022 W/m-K which is only 0.01% of bulk Si. We compared thermal conductivity of different Si nanostructures in Table I . It shows the thermal conductivity of 3D Si PnCs is even smaller than Si nanowires (NWs) and other Si nanostructures. Because PnCs do not have impurities and disorders, so the great reduction is due to the periodic spherical pores.

In bulk porous materials, the thermal conductivity value may be predicted from the Fourier classic Eucken model,[10,14] $\left(\kappa_{Porous}/\kappa_{Solid}\right)_{Eucken} = \left(1 - \phi\right)/\left(1 + \phi/2\right)$ , and Russell model,[15] $\left(\kappa_{Porous}/\kappa_{Solid}\right)_{Russell} = \left(1 - \phi^{2/3}\right)/\left(1 - \phi^{2/3} + \phi\right)$. When porosity is 90%, the predict value of $\kappa_{porous}$ is about 12 W/m-K and 11.2 W/m-K, which is even one order larger than the thermal conductivity of PnC with porosity of 30%. The classical Eucken model and Russell model is only good in structures which are much larger than the phonon mean free path. The models are not valid in nanoscale PnC, due to the strong size effect in nanostructures.[10,14]

Besides porosity effect, we also investigate the temperature effect on the thermal conductivity of Si PnCs in the range from 300 K to 1100 K (shown in Fig. 2(b)). Both the thermal conductivity of bulk Si and the thermal conductivity of Si PnCs with three different porosities, 50%, 80% and 90%, are calculated. Different from the bulk Si, it shows that the thermal conductivity of 3D PnCs change little as the increase of temperature. That is, the thermal conductivity of 3D PnCs is insensitive to temperature. A similar temperature dependence is also found in nanoporous SiGe[11] and thin Si NWs.[16] In pure bulk Si, there are normal phonon-phonon scatterings (crystal momentum conserved in three-phonon interactions) and Umklapp phonon-phonon scatterings (crystal momentum not conserved). The high-temperature of thermal conductivity is dominated by Umklapp scatterings and decrease as $\sim T^{-1}$. Different temperature dependence of thermal conductivity reflects different scattering mechanism.[16] In PnCs, there are plenty of boundary scatterings due to periodic spherical pores. Then the thermal conductivity will be independent on temperature when boundary scattering is the main mechanism.[17]

Participation ratio spectra of 3D Si PnCs with different porosities are shown in Fig. 3. We also calculated the participation ratio (P) of bulk Si, using a cell of 4×4×4 units[3] with periodic boundary condition. The values of P in bulk Si are almost in the range between 0.5 and 1, which represent extended phonon modes.[18] Obviously, the values of P in 3D PnCs are much smaller than those of bulk Si, which means phonon modes in PnCs are likely localized, that corresponds to the reduction of thermal conductivity in Si PnCs.

In order to quantitatively analyze the phonon localization in PnCs, we define the number of phonon modes whose value of P less than 0.5 divided by the total number of phonon modes as localization ratio (LR). According to the definition, a larger LR value means there are more localized phonon modes. Also shown in Fig. 3, LR values of PnCs with porosity of 70%, 80% and 90% are 34%, 58% and 91%, respectively. That is, there are more localized phonon modes with the increase of porosity. The extreme-low thermal conductivity of 3D PnCs mainly come from the localization of phonons.

To observe the spatial localization in 3D PhCs, we calculated the energy distribution. Only the modes whose P is lower than 0.2 are included in the summation of energy. (Calculation details in Supporting Information II) Fig. 1(c) and (d) show the normalized energy distribution on PnC at 300 K with porosity of 70% and 90%, respectively. It is obvious that the intensity of localized modes is much higher at boundaries. That is, most of phonons are localized at internal boundaries of PnCs. A heavier phonon localization in PnC with porosity of 90% corresponds to a lower thermal conductivity.

Using the Green-Kubo method, we have calculated thermal conductivities of nanoscale 3D Si PnCs with spherical pores. Results show that the thermal conductivity decreases as the increase of porosity. When the porosity is 30%, the thermal conductivity is 1.67 W/m-K. Moreover, when porosity increases to 90%, the thermal conductivity is decreased to 0.022 W/m-K which is only 0.01% of bulk Si. Due to the reduction in thermal conductivity, the PnC with spherical pores could lead to a larger ZT than unity. Moreover, we find that the thermal conductivity changes little when temperature increases from 300 K to 1100 K.

The participation ratio spectra of both PnCs and bulk Si have been calculated. It shows that the participation ratios of PnCs are smaller than those of bulk Si. That is, there are more phonons localized in PnCs at boundaries, which causes a lower thermal conductivity. Quantitatively, Si PnC with larger porosity has a larger LR value. As a consequence, the thermal conductivity of Si PnCs decreases as the increase of porosity.

In conclusion, an extreme-low thermal conductivity of nanoscale 3D Si PnC is predicted in this letter, which could lead to a larger ZT value than unity. There are much advances in obtaining the nanoscale 2D PnCs and 3D printing, which may help in fabricating nanoscale 3D PnCs in the future.

## Methods

In this work, the thermal conductivity of 3D PnCs is calculated by the Green-Kubo method, equilibrium molecular dynamics. EMD method is an effective method in calculating thermal conductivity in semiconductors.[4,19] We focus on the thermal conductivity of 3D PnCs at high temperature ($\geqslant$300 K). (Simulation details in the Supporting Information III)

To understand reduction of thermal conductivity by porosity, we carry out a vibration eigen-mode analysis on 3D Si PnCs. The mode localization can be qualitatively characterized by participation ratio P.[5] The participation ratio (P) for phonon mode k is defined through the normalized eigenvector $u_{i\alpha,k}$

$$P_k = \frac{1}{N \cdot \sum_{i=1}^{N} \left( \sum_{\alpha=1}^{3} u_{i\alpha,k}^* u_{i\alpha,k} \right)^2} \tag{1}$$

where $N$ is the total number of atoms, $u_{i,\alpha,k}$ is calculated by general utility lattice program (GULP).[20] When there are less atoms participating in the motion, the phonon mode has a smaller P value. For example, P is 1/$N$ when there is only one atom vibrates in the localized mode. When all atoms participate in the motion, P is calculated out as 1. That is, the smaller of the value of P the more localized of a phonon mode.

## Acknowledgements

This work was supported in part by the grant from the Asian Office of Aerospace R&D of the US Air Force (AOARD-114018) and MOE Grant R-144-000-305-112 (LY and BL), the startup fund from Tongji University (NY and BL) and the National Natural Science Foundation of China (Grant No. 11204216) (NY). NY is sponsored by Shanghai Pujiang program. NY is grateful to San-Huang Ke, Jun Shen, and Jin-Wu Jiang for useful discussions. LY is grateful to Jie Chen, Sha Liu, and Lifa Zhang for useful discussions. The authors thank the National Supercomputing Center in Tianjin (NSCC-TJ) for providing help in computations.

**Author contributions**

LY and NY carried out the numerical simulations and data analysis. BL and NY supervised the projects. All authors discussed the results and contributed to writing the manuscript.

**Additional information**

Competing financial interests: The authors declare no competing financial interests.

TABLE I. Comparison of thermal conductivity in different nanoscale materials.

| Material | Temperature [K] | Method | Thermal conductivity [W/m-K] |
|---|---|---|---|
| 3D Si PnC (porosity: 90%) | 300 | MD | 0.022 |
| Isotope-doped SiNW[3] | 300 | MD | 0.4 |
| Si nanotube[21] | 300 | MD | 3.0 |
| GeNWs with Si-coating[22] | 300 | MD | 4.7 |
| SiNWs with Ge-coating[23] | 300 | MD | 2.8 |
| Si/Ge superlattice[24] | 300 | MD | 1.2 |
| Nanoporous SiGe[11] | 300 | MD | 0.36 |
| 3D PnC (Ge QDs in Si)[8] | 300 | BTE | 0.95 |
| 2D Si PnC[7] | 300 | Experiment | 4.8 |
| Nanomesh[6] | 280 | Experiment | 1.9 |
| SiNW array[6] | 300 | Experiment | 3.5 |
| Bulk Si | 300 | MD | 170 |
| Bulk Si[12] | 300 | MD | 160 |
| Bulk Si[13] | 300 | Experiment | 156 |

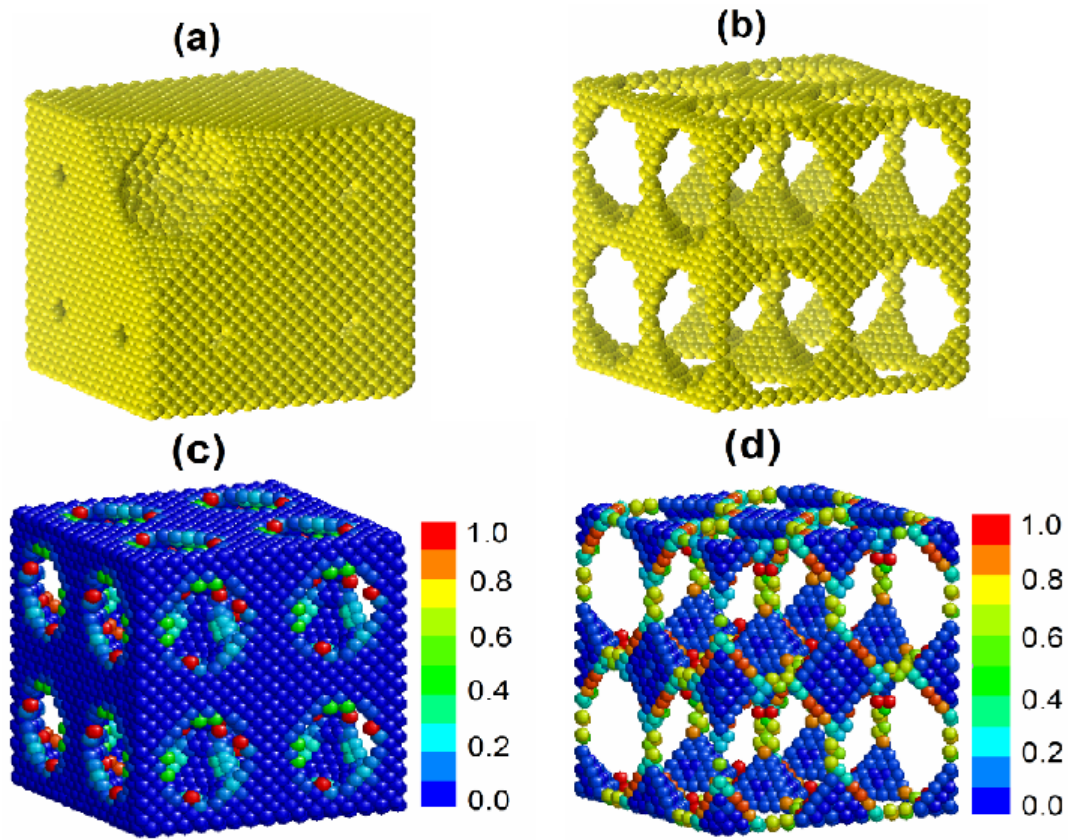

Fig. 1 The structures of nanoscale 3D Si PnCs. The period length of 3D PnCs is 8 units and the side length of simulation cell is 16 units. The periodic boundary condition is applied in simulation. The lattice constant is 0.543 nm of Si, and 1 unit represents 0.543 nm. (a) The structure of PnC with one corner cutting off and its porosity is 50%. (b) The structure PnC with porosity of 90%. (c) and (d) Normalized energy distribution on the PnC at 300 K with porosity of 70% and 90%, respectively. The intensity of the energy is depicted according to the color bar.

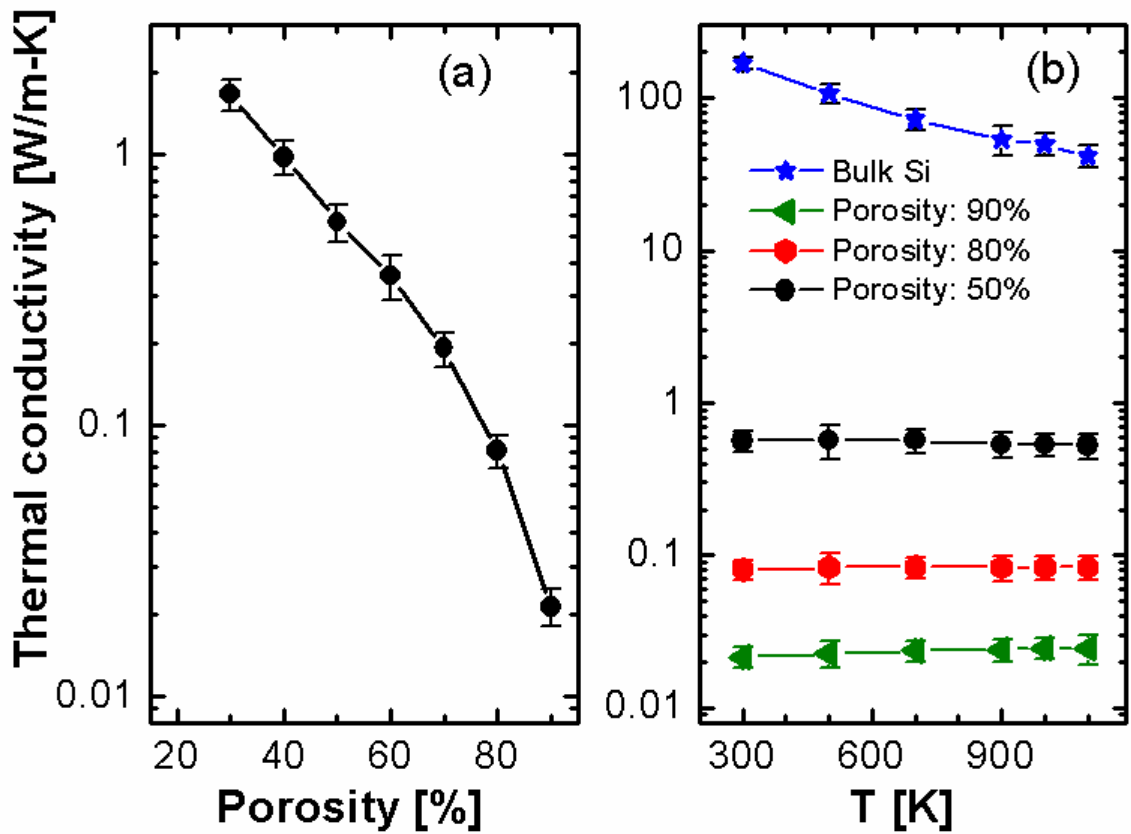

Fig. 2 (a) Thermal conductivity of PnCs versus porosity at 300 K. The side length of the simulation cell is 16 units and period length is 8 units. The thermal conductivity decreases rapidly as the increase of porosity. (b) The thermal conductivity of PnCs and bulk Si versus the temperature. The black dots, red hexagon and green triangle correspond to thermal conductivity of PnCs with porosity of 50%, 80% and 90%, respectively. Thermal conductivity of PnCs is insensitive to the temperature. The blue star is the thermal conductivity of bulk Si. The PnCs have the same period length (8 units) and side length (16 units). The cubic simulation cell of bulk Si has 12 units in side length. The lines are guides to the eye and the error bar is standard deviation of 12 simulations with different initial conditions.

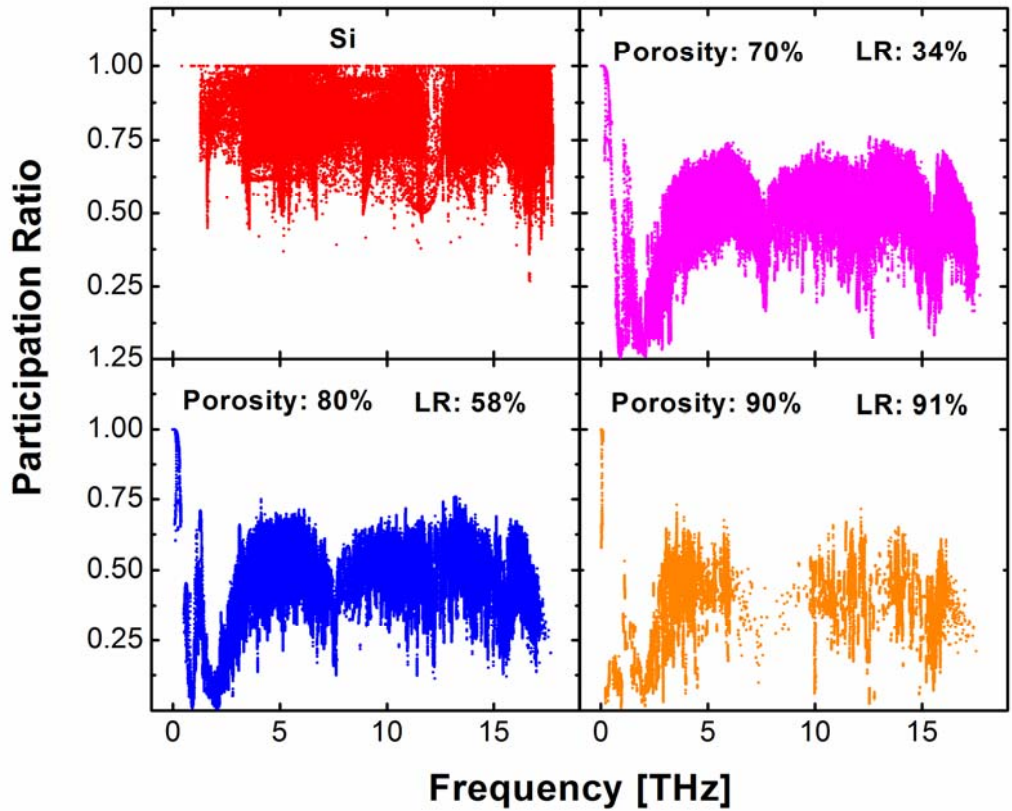

Fig. 3 The participation ratio spectra of Si PnCs and bulk Si. A cell of 4×4×4 units[3] is used in the calculation of participation ratio of bulk Si. Compared with bulk Si, the participation ratios of PnC have smaller value, which means phonon modes in PnC are likely localized. The participation ratio in bulk Si is almost in the range between 0.5 and 1.0 which characterizes the extended modes. LR values of PnCs are also shown in each channel. According to the definition of LR, there are more localized phonon modes in PnCs with larger porosity.

# Extreme Low Thermal Conductivity in Nanoscale 3D Si Phononic Crystal with Spherical Pores

Lina Yang[1], Nuo Yang[2*], and Baowen Li[1,2,3,4*]

[1]Department of Physics and Centre for Computational Science and Engineering, National University of Singapore, Singapore 117542, Republic of Singapore

[2]Center for Phononics and Thermal Energy Science, School of Physical Science and Engineering, Tongji University, 200092 Shanghai, People's Republic of China

[3]Graphene Research Center, National University of Singapore, Singapore 117542, Republic of Singapore

[4]NUS Graduate School for Integrative Sciences and Engineering, National University of Singapore, Singapore 117456, Republic of Singapore

*Correspondence and requests for materials should be addressed to: N.Y. (imyangnuo@gmail.com) and B.L. (phylibw@nus.edu.sg)

## I. Finite size effect in simulations

When using Green Kubo formula to calculate thermal conductivity, the finite size effect could arise if the simulation cell is not sufficiently large.[1] As shown in Table SI, we calculated thermal conductivity of PnCs with porosity of 90%, which have different sizes by EMD method at room temperature. The values of thermal conductivity change little when the side length of simulation cell is larger than 8 units. It shows that our simulation cell is large enough to overcome the finite size effect on calculating thermal conductivity. In all of the simulations of PnCs, we use 16 units as the side length of simulation cell and 8 units as period length.

## II. Energy distributions

To provide the information about the spatial distribution of a specific mode, local vibration density of states are defined as [2]

$$D_i(\omega) = \sum_k \sum_\alpha u^*_{i\alpha,k} u_{i\alpha,k} \delta(\omega - \omega_k)$$

(S1)

where $u_{i\alpha,k}$ is the $\alpha^{th}$ normalized eigenvector component of eigenmode k for atom i.

Based on the local vibration density of states, the spatial distribution of energy is calculated as

$$E_i = \sum_\omega (n + \frac{1}{2}) \hbar \omega D_i(\omega) = \sum_\omega \sum_k \sum_\alpha (n + \frac{1}{2}) \hbar \omega u^*_{i\alpha,k} u_{i\alpha,k} \delta(\omega - \omega_k)$$

(S2)

where n is the phonon occupation number given by the Bose-Einstein distribution. $E_i$ gives the total energy spatial distribution within a specified frequency regime, which provides a clear evidence of phonon localizations.

## III. MD simulation details

In simulations, the periodic boundary condition is applied in all three directions. To describe the interaction between Si atoms, we use Stillinger-Weber (SW) potential for Si,[1] which includes both two-body and three-body potential terms. The SW potential has been used widely to study the thermal properties of Si structures[3] for its accurate fit for experimental results on the thermal expansion coefficients. The heat current is defined as

$$\vec{J}_l(t) = \sum_i \vec{v}_i \varepsilon_i + \frac{1}{2} \sum_{ij} \sum_{i \neq j} \vec{r}_{ij} \left( \vec{F}_{ij} \cdot \vec{v}_i \right) + \sum_{ijk} \vec{r}_{ij} \left( \vec{F}_j(ijk) \cdot \vec{v}_j \right)$$

(S3)

where $\vec{F}_{ij}$ and $\vec{F}_{ijk}$ denote the two-body and three-body force, respectively. Thermal conductivity is derived from the Green-Kubo formula [1]

$$\kappa = \frac{1}{3 k_B T^2 V} \int_0^\infty < \vec{J}(\tau) \cdot \vec{J}(0) > d\tau$$

(S4)

where $\kappa$ is thermal conductivity, $k_B$ is the Boltzmann constant, V is the system volume defined as the cube of side length of the simulation cell, T is the temperature, and the angular bracket denotes an ensemble average.

Generally, the temperature in MD simulation, $T_{MD}$, is calculated from the kinetic energy of atoms according to the Boltzmann distribution:

$$\langle E \rangle = \sum_{i=1}^{N} \frac{1}{2} m v_i^2 = \frac{3}{2} N k_B T_{MD}$$

(S5)

where $\langle E \rangle$ is the total kinetic energy, $v_i$ is the velocity, m is the atomic mass, N is the number of particles in the system, and $k_B$ is the Boltzmann constant.

Velocity Verlet algorithm is employed to integrate equations of motion, and the MD step time is 1.0 fs. Initially, langevin heat reservoir is used to equilibrate the system at 300 K for 1 ns. Then, microcanonical ensemble (NVE) MD runs for another 16.7 ns. Meanwhile, heat current is recorded at each step. Then, the thermal conductivity is calculated by the Green-Kubo formula. The thermal conductivity is the mean value of twelve realizations with different initial conditions.

Fig. S1(a) shows a typical normalized heat current autocorrelation function (HCACF) used in Green-Kubo formula to calculate thermal conductivity of PnCs, where the side length of simulation cell is 16 units, the period length is 8 units, the porosity is 90%, and the temperature is 300 K. The heat current autocorrelation curve decays rapidly at the first one picoseconds, then followed by a slower decay to zero within 30 ps approximately. Such a two-stage decayed HCACF has been found in studies of other materials.[4] The first stage corresponds to the contribution of short wave length phonons to thermal conductivity, while the second stage corresponds to the contribution of long wave phonons. Chen et al. demonstrated that a small nonzero correction term, corresponding long wavelength phonons, can improve the accuracy of the two-exponential-fitting of HCACF.[4]

Fig. S1(b) shows the thermal conductivity which is an integration of HCACF. The thermal conductivity converges to 0.022 W/m-K after around 30 ps, which is consistent with the decay of heat current autocorrelation. The error bar is standard deviation of 12 simulations with different initial conditions.

TABLE SI. Comparison of thermal conductivity for different size of simulation cell. The porosity is 90% and the period length is 8 units. The temperature is set as 300 K.

| Dimensions | Side length (nm) | Thermal conductivity [W/mK] |
| --- | --- | --- |
| 8×8×8 | 4.34 | $(2.1\pm0.1)\times10^{-2}$ |
| 16×16×16 | 8.69 | $(2.2\pm0.3)\times10^{-2}$ |
| 24×24×24 | 13.03 | $(2.1\pm0.2)\times10^{-2}$ |

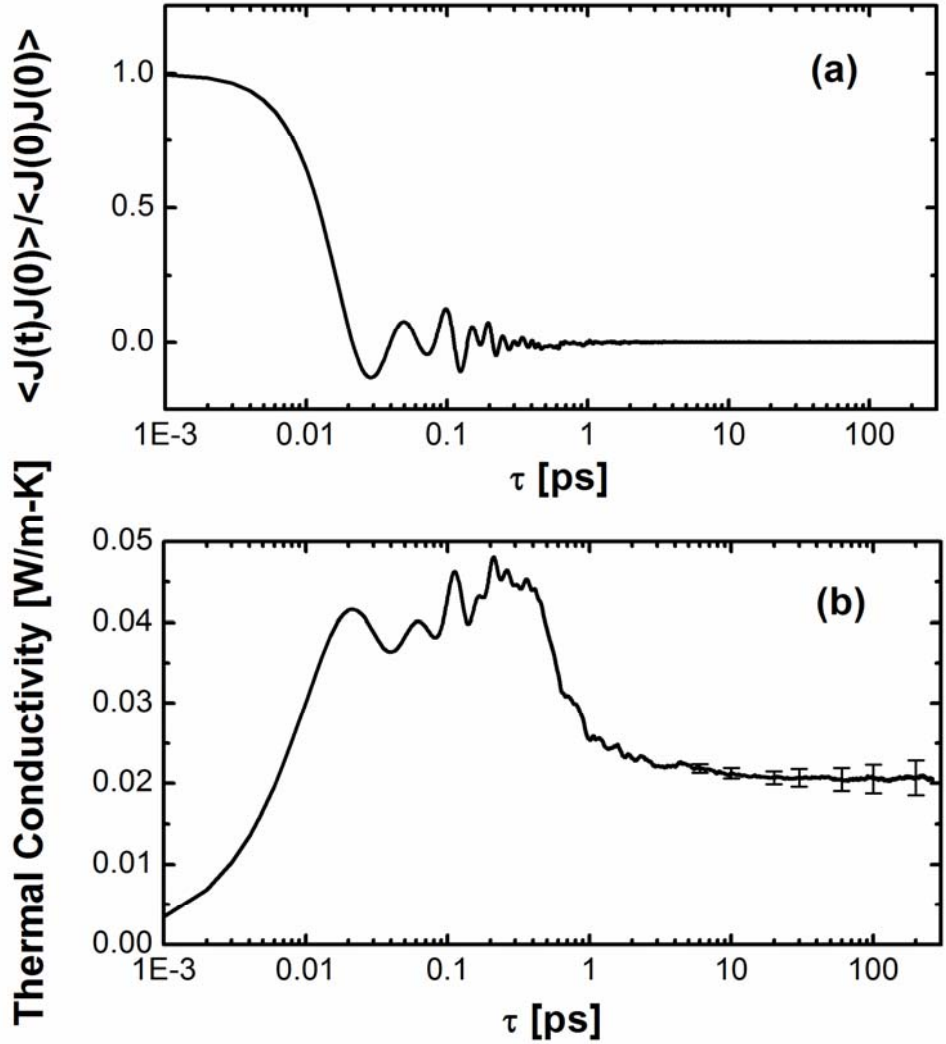

Fig. S1(a) Normalized heat current autocorrelation $\left\langle J(\tau) \cdot J(0) \right\rangle / \left\langle J(0) \cdot J(0) \right\rangle$ versus time τ for

PnC with porosity of 90% at 300 K. The side length of simulation cell is 16 units and the period length is 8 units. This figure shows heat flux correlation rapidly decays to zero in 30 ps. (b) Thermal conductivity calculated by integrating the correlation function in (a) versus time τ. The curve of thermal conductivity converges beyond 30 ps which consistent with the decay of heat current autocorrelation in (a).